# ELECTRONIC STATES IN A SUPERLATTICE CONSISTING OF ALTERNATING STRIPS OF SINGLE-LAYER AND BILAYER GRAPHENE

V.L. Abdrakhmanov[1], P.V. Badikova[1], D.V. Zav'yalov[1],
V. I. Konchenkov[1], S.V. Kryuchkov[1,2]

[1] 400005, Russia, Volgograd, Lenin Avenue 28, Volgograd State Technical University
[2] 400066, Russia, Volgograd, Lenin Avenue 27, Volgograd State Socio-Pedagogical University
polin.badicova@gmail.com

A model of a superlattice consisting of alternating strips of single-layer and bilayer graphene is proposed, whose parameters of the energy spectrum can be controlled by changing the external electric field perpendicular to the surface of the sample. Using the Kronig-Penney model, the dispersion equation is obtained, based on the analysis of which the energy spectrum of a graphene superlattice is studied depending on the ratio of the strip widths of single-layer and bilayer graphene. The results of the analytical solution are compared with the results of modeling by methods of the theory of the density functional. It is shown that the low-energy approximation used to derive the dispersion equation is valid when considering a superlattice with narrow strips of bilayer graphene and wide strips of single-layer graphene.

Introduction

Graphene superlattices (GSL) are in the focus of modern theoretical and experimental studies of low-dimensional semiconductor structures due to a number of interesting effects expected for such superlattices (SL), as well as in connection with the prospects of creating devices based on them for generating and detecting terahertz waves [1-3]. A characteristic feature of the GSL energy spectrum, as well as the graphene energy spectrum, is its non-additivity, which leads to an interdependence of current carrier motions in directions perpendicular to each other [4-6]. In addition to the effects expected for GSL by analogy with other types of SL, there have been recently discovered effects that are not manifested in other materials, but due to the two-dimensional nature of the GSL. So, in a recent work [7] it was shown that in a bilayer graphene, whose layers are rotated relative to each other at a certain angle, as a result of which a two-dimensional periodic potential is formed and the so-called moiré superlattice is formed, a transition to the superconducting state is observed. It is noted that such a structure can be used as a testing ground for studying high-temperature superconductivity.

Historically, the first to be studied were SL obtained by applying an external electric potential to graphene [8-12], as well as SL formed in a graphene sheet placed on a substrate of alternating strips of various dielectrics [13-20]. Also known SL formed in graphene interacting with a substrate having a periodic structure (self-organized array of periodically arranged metal nanospheres [21], nanostructured gold surface [22], corrugated rhodium foil surface [23], periodically located holes in the dielectric [24]). Widely studied moiré superlattices whose periodic potential arises due to the rotation of graphene layers (or other single-layer material) relative to each other [25-28]. In a number of works, much attention is paid to the appearance of additional Dirac points and lines — points in the reciprocal space of the crystal lattice, in which the valence and conduction bands intersect, and charge carriers can be described like relativistic massless particles [9, 12, 14, 16, 28]. Another important aspect is the existence and influence on the band structure of the formed material of surface states at the boundaries of various graphene modifications [14, 18] that make up the SL. Due to the periodicity of localization of surface states, they form a zone of allowed energy values, which may have a different shape than the zones formed by the main periodic potential [29 - 32]. A number of papers discuss the possibility of controlling the quantitative and qualitative parameters of the electronic energy spectrum of the GSL using external fields [10, 24, 33–35]; in [36], a similar situation is studied in a single-layer modification of silicon

— silicene. In this regard, it seems relevant to study electronic states in the GSL, which consists of alternating strips of single-layer and bilayer graphene, since it is known that due to the presence of a transverse electric field in the energy spectrum of bilayer graphene, a band gap can be formed. [10, 37]. In addition, the presence of alternation of single-layer and bilayer sections should lead to the formation of an additional periodic potential, which can manifest itself in the energy spectrum of SL. A theoretical study of the effect of a periodic in space constant electric field on the properties of bilayer graphene discussed in [10]. In [24] describes the manufacture and study of the transport properties of SL formed in single-layer graphene placed over a substrate with a two-dimensional periodic structure of alternating holes in a dielectric under the influence of a constant electric field. The work [38] is devoted to studying the electronic energy spectrum of SL from alternating strips of single-layer and bilayer graphene without taking into account the effect of the transverse electric field on the region of bilayer graphene.

In this work, an attempt is made to consider the electronic states in the GSL, which consists of alternating strips of single-layer and bilayer graphene, placed in a constant electric field perpendicular to the surface of the sample. Figure 1 shows the location of additional graphene strips forming bilayer regions. It is assumed that the bilayer regions are of type AB (the so-called Bernal stacking), the arrangement of atoms of the second layer of graphene relative to the first layer is shown in the insert. In Section 1, the energy spectrum of the considered GSL is studied analytically, and in Section 2, the results of quantum-chemical modeling by the methods of the density functional theory are presented.

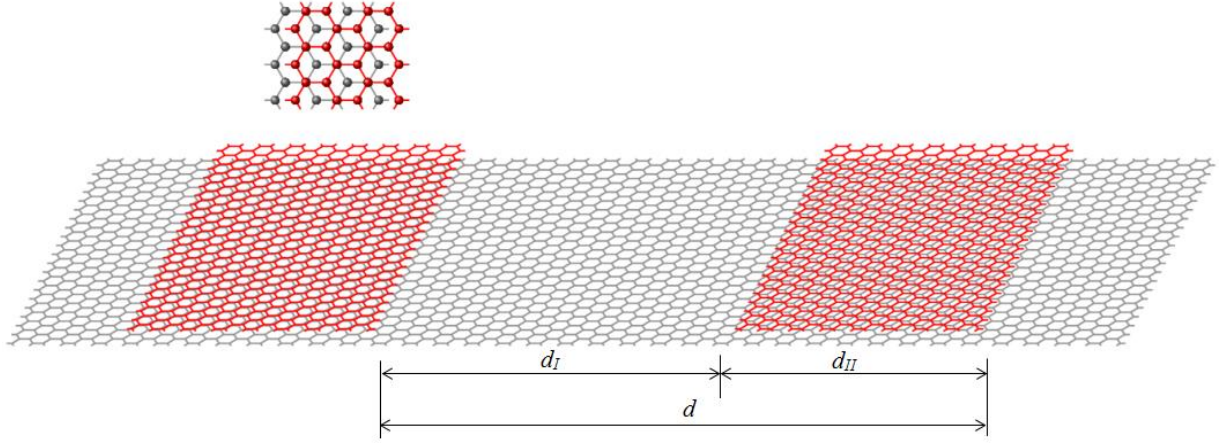

Figure 1 - Layout of graphene strips forming bilayer regions on the main graphene sheet.

1. Derivation of the dispersion equation for a superlattice consisting of alternating strips of single-layer and bilayer graphene

The energy spectrum of bilayer graphene, to which a transverse constant electric field is applied (the so-called bias bilayer graphene), is determined by the following expression [37]:

$$\varepsilon_\alpha = \pm\sqrt{\Delta^2 + h^2 v_F^2 k^2 + \frac{t_\perp^2}{2} \pm \sqrt{h^2 v_F^2 k^2 \left(4\Delta^2 + t_\perp^2\right) + \frac{t_\perp^4}{4}}}. \qquad (1)$$

Here $v_F \sim 10^8\, cm/c$ is the velocity on the Fermi surface in graphene, $h\mathbf{k}$ is the quasimomentum of the electron, $t_\perp \approx 0.4\, eV$ is the overlap integral between the layers of two-layer graphene, $\Delta \sim 0.1...0.3 t_\perp$ is a parameter that determines the band gap (half-width of the band gap $\Delta' = \Delta/\sqrt{4(\Delta/t_\perp)^2 + 1} \approx \Delta$).

To obtain the dispersion relation, we use the approach developed in [13, 38], which is a modified Kronig – Penney model.

The Hamiltonian of bilayer graphene has the form:

$$H_{II} = \begin{pmatrix} \Delta & \hbar v_F\left(-i\dfrac{\partial}{\partial x} - \dfrac{\partial}{\partial y}\right) & t_\perp & 0 \\ \hbar v_F\left(-i\dfrac{\partial}{\partial x} + \dfrac{\partial}{\partial y}\right) & \Delta & 0 & 0 \\ t_\perp & 0 & -\Delta & \hbar v_F\left(-i\dfrac{\partial}{\partial x} + \dfrac{\partial}{\partial y}\right) \\ 0 & 0 & \hbar v_F\left(-i\dfrac{\partial}{\partial x} - \dfrac{\partial}{\partial y}\right) & -\Delta \end{pmatrix} \qquad (2)$$

The solution to the Schrödinger equation is a four-component spinor $\Psi^{II} = (\psi_1, \psi_2, \psi_3, \psi_4)^T$. Components $\psi_1, \psi_2$ correspond to states on the sublattices $A_1$ and $B_1$ of one sheet, and components $\psi_3, \psi_4$ correspond to states on the sublattices $A_2$ and $B_2$ of another sheet of graphene. In general, the Hamiltonian is written near one of the Dirac points. Since in order to obtain a translation matrix, it is necessary to fulfill the matching conditions for the wave function [13]:

$$\Psi_n^I(d_I - 0) = \Psi_n^{II}(d_I + 0), \quad \Psi_n^{II}(d - 0) = \Psi_{n+1}^I(+0). \qquad (3)$$

So we have to write the Hamiltonian of single-layer graphene also in the form of a matrix of size $4 \times 4$. Formally, this can be done by setting $t_\perp = \Delta = 0$ in (2) — in this case, the eigenvalues of the Hamiltonian will be equal to $E_I = \pm v_F \hbar \sqrt{k_x^2 + k_y^2}$, which corresponds to the spectrum of single-layer graphene in the low-energy approximation (this is the approach used in [38], where a GSL consisting of alternating strips of a single-layer and bilayer graphene (unbiased bilayer graphene), for which $\Delta = 0$ is assumed).

On the other hand, with this approach, the components of the spinor, which is the solution of the Schrödinger equation $H_I \Psi^I = E_I \Psi^I$, have a different meaning: $\psi_1, \psi_2$ correspond to the states on the sublattices $A_1$ and $B_1$ for the electron near the Dirac point $K_1$, $\psi_3, \psi_4$ to the states on the same sublattices, but corresponding to the Dirac point $K_2$.

In the low-energy approximation, in the absence of electron transfer between the valleys, these states are independent of each other. It is the independence of states that allows one to count the values of the quasimomentum from the nearest Dirac point, which makes it possible to expand the two-line matrix of the Hamiltonian of single-layer graphene to four-line by simply adding a complex conjugate matrix as the second block of a block-diagonal matrix of size $4\times 4$. So, the Hamiltonian of single-layer graphene can be written as

$$H_I = \begin{pmatrix} 0 & hv_F\left(-i\dfrac{\partial}{\partial x}-\dfrac{\partial}{\partial y}\right) & 0 & 0 \\ hv_F\left(-i\dfrac{\partial}{\partial x}+\dfrac{\partial}{\partial y}\right) & 0 & 0 & 0 \\ 0 & 0 & 0 & hv_F\left(-i\dfrac{\partial}{\partial x}+\dfrac{\partial}{\partial y}\right) \\ 0 & 0 & hv_F\left(-i\dfrac{\partial}{\partial x}-\dfrac{\partial}{\partial y}\right) & 0 \end{pmatrix} \quad (4)$$

The solution to the Schrödinger equation in the well region will be sought in the form:

$$\Psi^I = A\psi^I \exp(ik_y y) = A\left(\begin{pmatrix} \varphi_{1p} \\ \varphi_{2p} \\ \varphi_{3p} \\ \varphi_{4p} \end{pmatrix}\exp(ik_1 x) + \begin{pmatrix} \varphi_{1n} \\ \varphi_{2n} \\ \varphi_{3n} \\ \varphi_{4n} \end{pmatrix}\exp(-ik_1 x)\right)\exp(ik_y y). \quad (5)$$

Substituting (5) into (4), we obtain the following expression for the wave function

$$\psi^I = \underbrace{\begin{pmatrix} 1 & 1 & 0 & 0 \\ \dfrac{\lambda_+}{E} & -\dfrac{\lambda_-}{E} & 0 & 0 \\ 0 & 0 & 1 & 1 \\ 0 & 0 & \dfrac{\lambda_-}{E} & -\dfrac{\lambda_+}{E} \end{pmatrix}}_{G}\begin{pmatrix} e^{ik_1 x} & 0 & 0 & 0 \\ 0 & e^{-ik_1 x} & 0 & 0 \\ 0 & 0 & e^{ik_1 x} & 0 \\ 0 & 0 & 0 & e^{-ik_1 x} \end{pmatrix}\begin{pmatrix} \varphi_{1p} \\ \varphi_{1n} \\ \varphi_{3p} \\ \varphi_{3n} \end{pmatrix} \quad (6)$$

Here we use the designations $\lambda_\pm = h v_F (k_1 \pm i k_y)$, the value of $k_1$ can be found from condition $E^2 = h^2 v_F^2 (k_1^2 + k_y^2)$. It is seen from (6) that the pairs of components $\varphi_{1p}$, $\varphi_{1n}$ and $\varphi_{3p}$, $\varphi_{3n}$ are independent of each other.

In the barrier region, two situations need to be considered: when the electron energy is less than $\Delta$ – in this case the solution is damping $\psi^{II} = B_1 \exp(k_2 x) + B_2 \exp(-k_2 x)$, and the case when the energy is higher than $\Delta$ - in this case the solution is oscillating: $\psi^{II} = C_1 \exp(i k_2 x) + C_2 \exp(-i k_2 x)$.

In the first case, we have

$$\psi^{II} = \begin{pmatrix} 1 & 1 & 0 & 0 \\ \dfrac{\varkappa_-}{\Delta - E} & -\dfrac{\varkappa_+}{\Delta - E} & 0 & 0 \\ 0 & 0 & b & b \\ 0 & 0 & -\dfrac{\varkappa_+ b}{\Delta + E} & \dfrac{\varkappa_- b}{\Delta + E} \end{pmatrix} \underbrace{\begin{pmatrix} e^{k_2 x} & 0 & 0 & 0 \\ 0 & e^{-k_2 x} & 0 & 0 \\ 0 & 0 & e^{k_2 x} & 0 \\ 0 & 0 & 0 & e^{-k_2 x} \end{pmatrix}}_{J} \begin{pmatrix} \psi_{1p} \\ \psi_{1n} \\ \psi_{1p} \\ \psi_{1n} \end{pmatrix}, \quad (7)$$

$b = -\dfrac{(\Delta - E)^2 - \varkappa_+ \varkappa_-}{(\Delta - E) t_\perp}$, $\varkappa_\pm = i h v_F (k_2 \pm k_y)$. This expression is valid under the condition

$$(4\Delta^2 + t_\perp^2) E^2 - \Delta^2 t_\perp^2 > 0, \quad (8)$$

as well as conditions

$$k_y^2 - \dfrac{(\Delta^2 + E^2)}{h^2 v_F^2} + \dfrac{\sqrt{(4\Delta^2 + t_\perp^2) E^2 - \Delta^2 t_\perp^2}}{h^2 v_F^2} \geq 0, \quad (9)$$

while $k_2 = k_2^+ = \sqrt{k_y^2 - \dfrac{(\Delta^2 + E^2)}{h^2 v_F^2} + \dfrac{\sqrt{(4\Delta^2 + t_\perp^2) E^2 - \Delta^2 t_\perp^2}}{h^2 v_F^2}}$ or the condition

$$k_y^2 - \dfrac{(\Delta^2 + E^2)}{h^2 v_F^2} - \dfrac{\sqrt{(4\Delta^2 + t_\perp^2) E^2 - \Delta^2 t_\perp^2}}{h^2 v_F^2} \geq 0, \quad (10)$$

while $k_2 = k_2^- = \sqrt{k_y^2 - \dfrac{(\Delta^2 + E^2)}{\hbar^2 v_F^2} - \dfrac{\sqrt{(4\Delta^2 + t_\perp^2)E^2 - \Delta^2 t_\perp^2}}{\hbar^2 v_F^2}}$ .

An oscillating solution can be obtained from (7) by a formal replacement of $k_2 \to i\kappa_2$.

To derive the dispersion relation, it is required to find the transfer matrix ($T$-matrix), which relates the values of the spinor components for the $l$-th supercell to the spinor components for the $(l + 1)$-th supercell:

$$\begin{pmatrix} \varphi_{1p}(l+1) \\ \varphi_{2p}(l+1) \\ \varphi_{3p}(l+1) \\ \varphi_{4p}(l+1) \end{pmatrix} = T \begin{pmatrix} \varphi_{1p}(l) \\ \varphi_{2p}(l) \\ \varphi_{3p}(l) \\ \varphi_{4p}(l) \end{pmatrix}. \quad (11)$$

To derive the dispersion equation, we use the technique described in detail in [13], we obtain in the case of a decaying solution in the barrier region.

$$T = G^{-1} J \begin{pmatrix} e^{k_2 d_{II}} & 0 & 0 & 0 \\ 0 & e^{-k_2 d_{II}} & 0 & 0 \\ 0 & 0 & e^{k_2 d_{II}} & 0 \\ 0 & 0 & 0 & e^{-k_2 d_{II}} \end{pmatrix} J^{-1} G \begin{pmatrix} e^{ik_1 d_I} & 0 & 0 & 0 \\ 0 & e^{-ik_1 d_I} & 0 & 0 \\ 0 & 0 & e^{ik_1 d_I} & 0 \\ 0 & 0 & 0 & e^{-ik_1 d_I} \end{pmatrix}. \quad (12)$$

According to [13, 39], the dispersion relation is determined by the expression

$$Tr T = 4\cos k_x d, \quad (13)$$

where $Tr T$ is the trace of the T-matrix (the sum of the elements standing on the main diagonal).

The dispersion relation, which allows us to calculate the electron energy $E_n(k_x, k_y)$ ($n$ is the miniband number, $k_x, k_y$ are the components of the electron quasimomentum), can be represented as:

$$F(E, k_1(E), k_2(E), k_x, k_y) = 0. \quad (14)$$

The calculation shows that for the considered GSL there are two families of dispersion surfaces corresponding to the presence of two branches in the spectrum of bilayer graphene. In addition to the situations considered above, when $k_1$ is a real number, formally there is a solution of the dispersion equation for imaginary values of $k_1$, which indicates the formation of the so-called Tamm minibands that arise due to the periodic arrangement of surface states at the boundaries of single-layer and bilayer graphene regions.

Let us turn to dimensionless variables: $q_{x,y} = k_{x,y}d$, $q_{1,2} = k_{1,2}d$, $a_I = d_I/d$, $a_{II} = d_{II}/d$, $B = \hbar v_F/(t_\perp d)$. The values of $E$ and $\Delta$ will be measured in units of $t_\perp$. We introduce the following notation:

$$Q_2^+ = \left(B^2 q_y^2 - (\Delta^2 + E^2) + \sqrt{(4\Delta^2 + 1)E^2 - \Delta^2}\right)/B^2, \quad (15)$$

$$Q_2^- = \left(B^2 q_y^2 - (\Delta^2 + E^2) - \sqrt{(4\Delta^2 + 1)E^2 - \Delta^2}\right)/B^2, \quad (16)$$

$$Q_1 = (E^2 - B^2 q_y^2)/B^2. \quad (17)$$

$$F_1 = 2\cos q_1 a_I \operatorname{ch} q_2 a_{II} + \frac{E^2(q_1^2 - q_2^2) + \Delta^2(q_y^2 - q_1^2)}{(\Delta^2 - E^2)q_1 q_2} \sin q_1 a_I \operatorname{sh} q_2 a_{II} - 2\cos q_x \quad (18)$$

$$F_2 = 2\cos q_1 a_I \cos q_2 a_{II} + \frac{E^2(q_1^2 + q_2^2) + \Delta^2(q_y^2 - q_1^2)}{(\Delta^2 - E^2)q_1 q_2} \sin q_1 a_I \sin q_2 a_{II} - 2\cos q_x \quad (19)$$

$$F_3 = 2\operatorname{ch} q_1 a_I \operatorname{ch} q_2 a_{II} + \frac{-E^2(q_1^2 + q_2^2) + \Delta^2(q_y^2 + q_1^2)}{(\Delta^2 - E^2)q_1 q_2} \operatorname{sh} q_1 a_I \operatorname{sh} q_2 a_{II} - 2\cos q_x \quad (20)$$

$$F_4 = 2\operatorname{ch} q_1 a_I \cos q_2 a_{II} + \frac{-E^2(q_1^2 - q_2^2) + \Delta^2(q_y^2 + q_1^2)}{(\Delta^2 - E^2)q_1 q_2} \operatorname{sh} q_1 a_I \sin q_2 a_{II} - 2\cos q_x \quad (21)$$

The first family of surfaces can be obtained by solving the dispersion equation (14) with the following parameter values:

$$Q_1 \geq 0,\ Q_2^+ \geq 0:\ q_1 = \sqrt{Q_1},\ q_2 = \sqrt{Q_2^+},\ F = F_1;$$

$$Q_1 \geq 0,\ Q_2^+ < 0:\ q_1 = \sqrt{Q_1},\ q_2 = \sqrt{-Q_2^+},\ F = F_2;$$

$$Q_1 < 0,\ Q_2^+ \geq 0:\ q_1 = \sqrt{-Q_1},\ q_2 = \sqrt{Q_2^+},\ F = F_3;$$

$$Q_1 < 0,\ Q_2^+ < 0:\ q_1 = \sqrt{-Q_1},\ q_2 = \sqrt{-Q_2^+},\ F = F_4.$$

(22)

The second family:

$$Q_1 \geq 0,\ Q_2^- \geq 0:\ q_1 = \sqrt{Q_1},\ q_2 = \sqrt{Q_2^-},\ F = F_1;$$

$$Q_1 \geq 0,\ Q_2^- < 0:\ q_1 = \sqrt{Q_1},\ q_2 = \sqrt{-Q_2^-},\ F = F_2;$$

$$Q_1 < 0,\ Q_2^- \geq 0:\ q_1 = \sqrt{-Q_1},\ q_2 = \sqrt{Q_2^-},\ F = F_3;$$

$$Q_1 < 0,\ Q_2^- < 0:\ q_1 = \sqrt{-Q_1},\ q_2 = \sqrt{-Q_2^-},\ F = F_4.$$

(23)

The dispersion equations were solved numerically by the Newton method.

Figure 2 shows the four bottom minibands, the arrangement of which corresponds to the conduction band of the basic material (the case is considered when the strip width of bilayer graphene is much larger than the strip width of single-layer graphene ($a_I = 0.05$, $a_{II} = 0.95$), value $\Delta = 0.1$ in units of $t_\perp$). The two lower minibands are determined by the solution of the dispersion equation in the form (23), the two upper minibands in the form (22). It is seen that in the situation under consideration, energy gaps appearing between minibands are significant, i.e., the structure under consideration must have the properties of a semiconductor. The energy spectrum is periodic on $q_x$.

It should be noted that the shape of the dispersion surfaces corresponding to different families differs significantly in the case of wide strips of bilayer and narrow strips of single-layer graphene. In the opposite case, the considered families of dispersion surfaces have a similar shape, but are spaced a small distance in energy. Figure 3 shows the dispersion lines constructed for case $a_I = 0.95$, $a_{II} = 0.05$ and $\Delta = 0$. The solid line indicates the dispersion curves described by expressions (23), the dashed line indicates the curves described by expression (22).

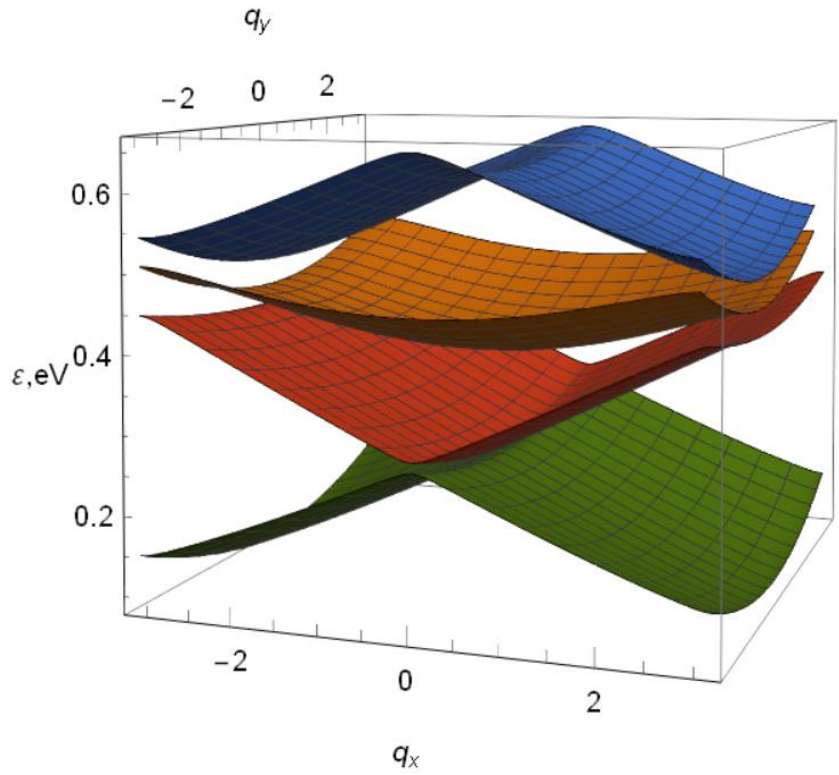

Figure 2 – Dispersion surfaces of charge carriers at $a_I = 0.05$; $a_{II} = 0.95$; $\Delta = 0.1$

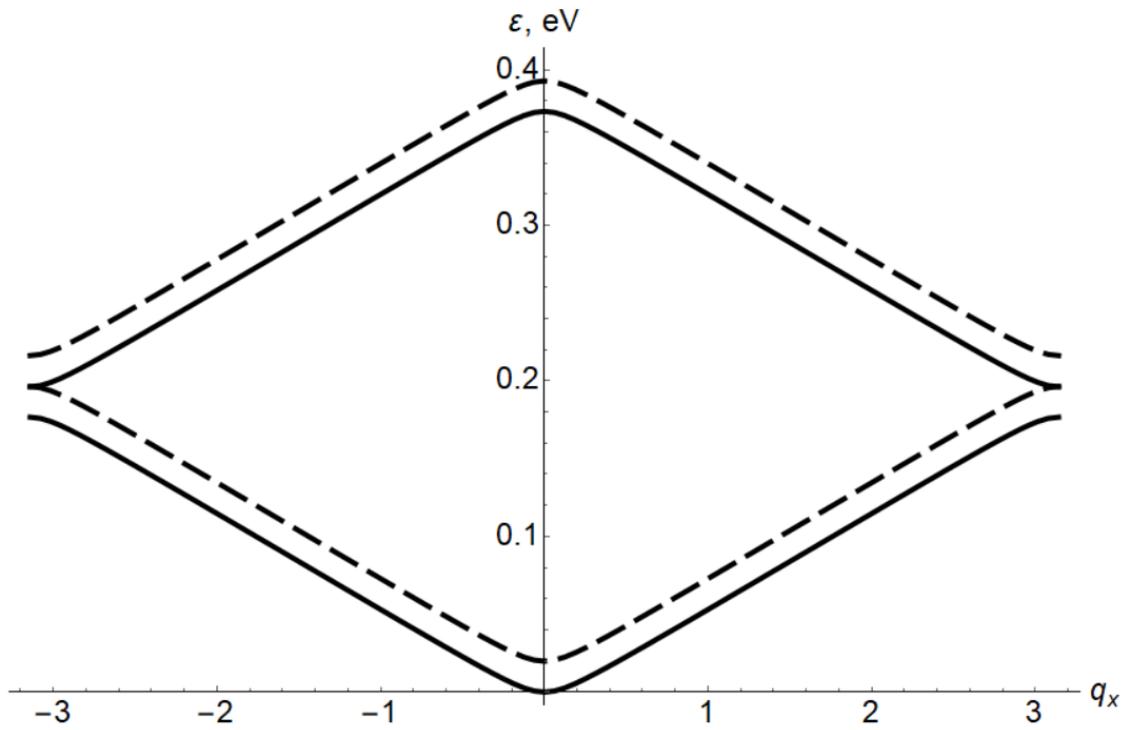

Figure 3 – Dispersion curves plotted for case $\Delta = 0$, $q_y = 0$, $a_I = 0.95$, $a_{II} = 0.05$.

It can be seen that the upper miniband of the family described by expressions (23) intersects with the lower miniband of the family described by expression (22) at values of $q_x = \pm \pi$, i.e., the so-called Dirac lines are present in the energy spectrum of the studied GSL. Figure 4 shows the dependence of the electron energy on $q_y$ under condition $q_x = \pi$ for such minibands at different values of the width of the wells and barriers. The calculation shows that, at close values of $a_I$ and $a_{II}$, the energy gap is maximum, in the case of narrow barriers, the energy gap tends to zero, which corresponds to Dirac lines. At $\Delta \to 0$, the minimum value of energy in the lowest mini-band described by expressions (23), located at point $\{q_x, q_y\} = \{0, 0\}$, is zero, that is, the conduction band and the valence band of the considered GSL are touched at this point, which is a characteristic feature of single-layer graphene and bilayer graphene in lack of an external transverse electric field.

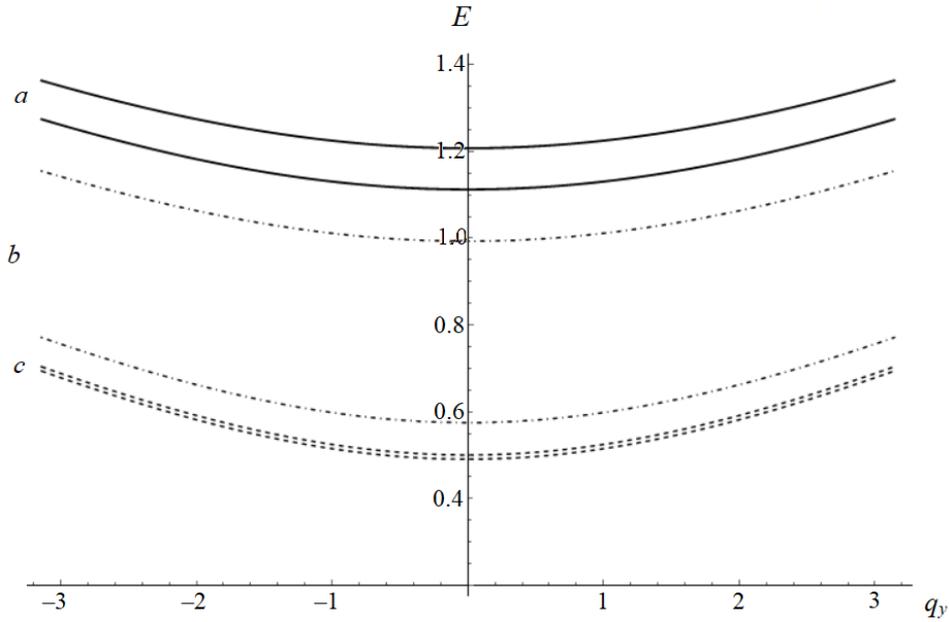

Figure 4.– Dispersion curves plotted for case $\Delta = 0.1; k_x = \pi$ described by dispersion equation (22) in the case of different relative widths of the well and the barrier: a) $a_I = 0.05$, $a_{II} = 0.95$; b) $a_I = 0.5$, $a_{II} = 0.5$; c) $a_I = 0.99$, $a_{II} = 0.01$

Thus, using the *T*-matrix method, dispersion relations are obtained that describe two families of dispersion surfaces. The obtained GSL energy spectrum has a miniband character, and the width of the energy gaps between the minibands strongly depends on the ratio of the well width to the barrier. It is significant that along with the solutions of the Schrödinger equation oscillating in the well region, there are the decay solutions in the well region, corresponding to the Tamm states.

2. Quantum-chemical modeling of the energy spectrum

The study of the energy spectrum of the GSL considered above were conducted using the OpenMX package. OpenMX (Open source package for Material eXplorer) is a software package for nano-scale material simulations based on density functional theories (DFT), norm-conserving pseudopotentials, and pseudoatomic localized basis functions. [40 - 43]. This allows us to more efficiently calculate properties of periodic systems consisting of hundreds and thousands of atoms with less time costs than plane waves based ones. The methods and algorithms used in OpenMX and their implementation are carefully designed for the realization of large-scale ab initio electronic structure calculations on parallel computers. The calculations were performed for a SL with a period $d = 10.9$ nm, which corresponds to the 51 period of the main graphene lattice, with variable width of the second layer. During the SCF calculation, a PBE functional was used, and integration over the Brillouin zone was performed on a regular 4 × 16 grid. The accuracy of the calculation was $10^{-6}$ hartree.

The crystal lattice of graphene is a hexagonal lattice which constant is $a_0 = 2.46$ Å. The primitive graphene cell includes 2 atoms located in the plane of the lattice. The lattice parameters of single-layer graphene coincide with the parameters of the primitive lattice of bilayer graphene, consisting of 4 atoms [44]. We consider bilayer graphene of type AB (Bernal stacking graphene, see Figure 1). In both single-layer and bilayer graphene, the minimum of the conduction band is located at the K-point of the Brillouin zone. Strips of single-layer and bilayer graphene in the considered superlattice alternate in the direction of the X axis. The values of the quasi-wave vectors are counted from the K-point.

The spectrum of a GSL with different ratios of the strip widths of single-layer and bilayer graphene is studied. We consider, first of all, the limiting cases. Figure 5 shows the dispersion surfaces constructed for the case of bilayer graphene ($d_{II} = d$); the quasi-wave vectors are counted from the K-point. Near the bottom of the conduction band, there are 4 distinct singularities similar in shape to the Dirac points (Figure 5b). In the case of influence to a sample of a constant transverse electric field, an energy gap appears (Figure 5c), and the singularities are smoothed out (Figure 5d). The appearance of additional Dirac points located at the vertices and center of a regular triangle in the spectrum of AB bilayer graphene was noted in a number of works devoted to quantum-chemical modeling of this material (see, for example, review [44]).

Figure 6 shows the results of a study of the dependence of the shape of the dispersion surfaces on the width of the strip $d_{II}$ of bilayer graphene in relation to the period of the superlattice. Energy is measured in electron volts. It can be seen that in the limiting case $d_{II} = 0$ (in fact, single-layer graphene) near the K-point of basic material, the spectrum of the structure under consideration has the form of Dirac cones (Figure 6a).

Figures 6b) - 6f) show that with increasing strip width of bilayer graphene, the energy spectrum of the superlattice under consideration changes significantly (the values of the quasi-wave vectors are counted from the K point). The simulation showed, firstly, that the minibands forming the Dirac cone are shifted to the high-energy region by about 0.1 eV, and even in the absence of a transverse electric field, a band gap is formed. Secondly, the energy spectrum of the studied GSL is periodic in the direction of alternating layers and is even in $q_x$ (Figure 7), and this result is valid for all relative widths of stripes of single-layer and bilayer graphene. Third, the energy spectrum becomes asymmetric in $q_y$ as the ratio of the strip width of bilayer graphene to the period of the SL increases. This result can be explained in the following way. When the strip width of bilayer graphene is small compared to the SL period (Figures 6b, 6c), we can consider the carbon atoms in the second graphene layer as regularly located impurities of high concentration, which explains the energy shift of the position of the Dirac point. On the other hand, the periodicity of the location of such an "impurity" leads to the formation of minibands, which, due to the large difference between the superlattice period and the crystal lattice period of the base material (single-layer graphene), do not have to be symmetrical about the K point. As can be seen from Fig. 6b), 6c), 8, for a small relative strip width of bilayer graphene ($d_{II}/d = 3/51$), the asymmetry of the dispersion curves in $q_y$ is small, therefore, under such conditions, the low-energy approximation used in Section 1 for the analytical consideration of the properties of superlattices can be considered as proved.

Figure 9 shows the two minibands closest to the Dirac point related to the conduction band of the starting material, calculated using the methods of the density functional theory and based on the modified Kronig-Penney model described in Part 1. It can be seen that the shape of the dispersion surfaces calculated by different methods is similar.

An analytical consideration based on the solution of the Schrödinger equation does not take into account the displacement of the Fermi level, believing it to be the same for single-layer graphene and for a SL consisting of alternating strips of single-layer and bilayer graphene. Quantum-chemical calculations showed that the position of the Dirac point with the addition of an additional strip of graphene shifts to the high-energy region by about 0.1 eV, and even in the absence of a transverse electric field, a band gap is formed.

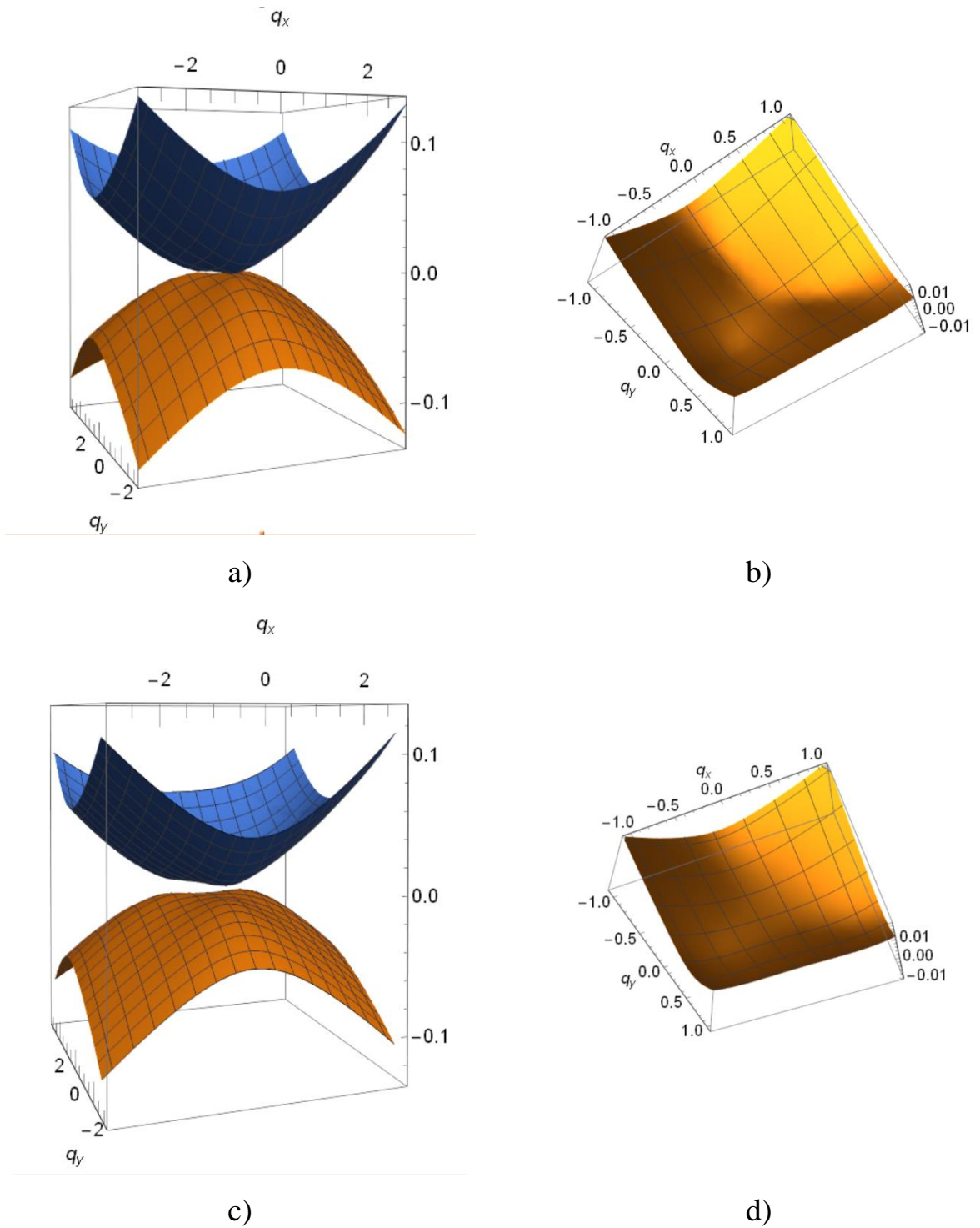

a)  b)

c)  d)

Figure 5. - Dispersion surfaces closest to the Fermi level obtained for the case of bilayer graphene. a), b) - in the absence of an external transverse electric field; c), d) - under the influence of a transverse electric field $10^6$ V / cm.

Upon analytical consideration at $\Delta = 0$, which corresponds to the absence of an external transverse electric field, the energy in the minimum of the lower miniband corresponding to the K-point is zero. Thus, quantum-chemical modeling showed that under condition $d_{II} \ll d$, that is, with a small strip width of bilayer graphene compared to the strip width of single-layer graphene in the considered superlattice, the use of the low-energy approximation in the Kronig-Penney model, which serves to derive the dispersion equation, is valid. The energy spectrum of a superlattice consisting of alternating strips of single-layer and bilayer graphene has a miniband character.

It should be noted that under the assumption that the strips of bilayer graphene are narrower than the strips of single-layer graphene, the considered superlattice turns out to be ideologically close to one of the first GSL models proposed in [45]. In [45], a GSL formed in single-layer graphene as a result of the deposition of periodically arranged lines of hydrogen atoms on it was considered.

The GSL considered in this work has forbidden and allowed miniband widths of the order of hundredths of an electron-volt, and these parameters can be controlled by applying an external electric field perpendicular to the surface of the sample.

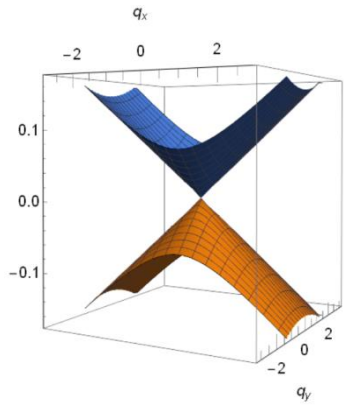

a) $d_{II}/d = 0$

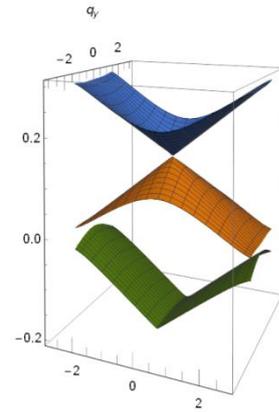

b) $d_{II}/d = 2/51$

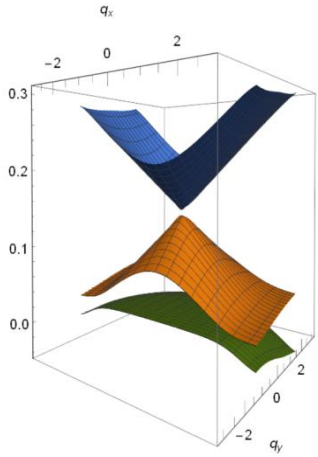

c) $d_{II}/d = 3/51$

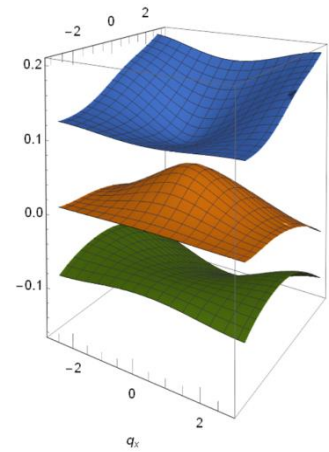

d) $d_{II}/d = 24/51$

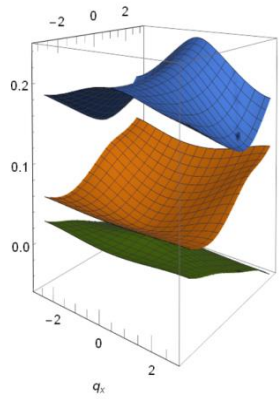

e) $d_{II}/d = 48/51$

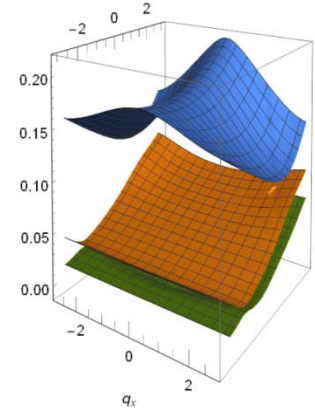

f) $d_{II}/d = 50/51$

Figure 6. - Dependence of the shape of dispersion surfaces near the Dirac point on the strip width of bilayer graphene in the absence of an external electric field perpendicular to the sample surface.

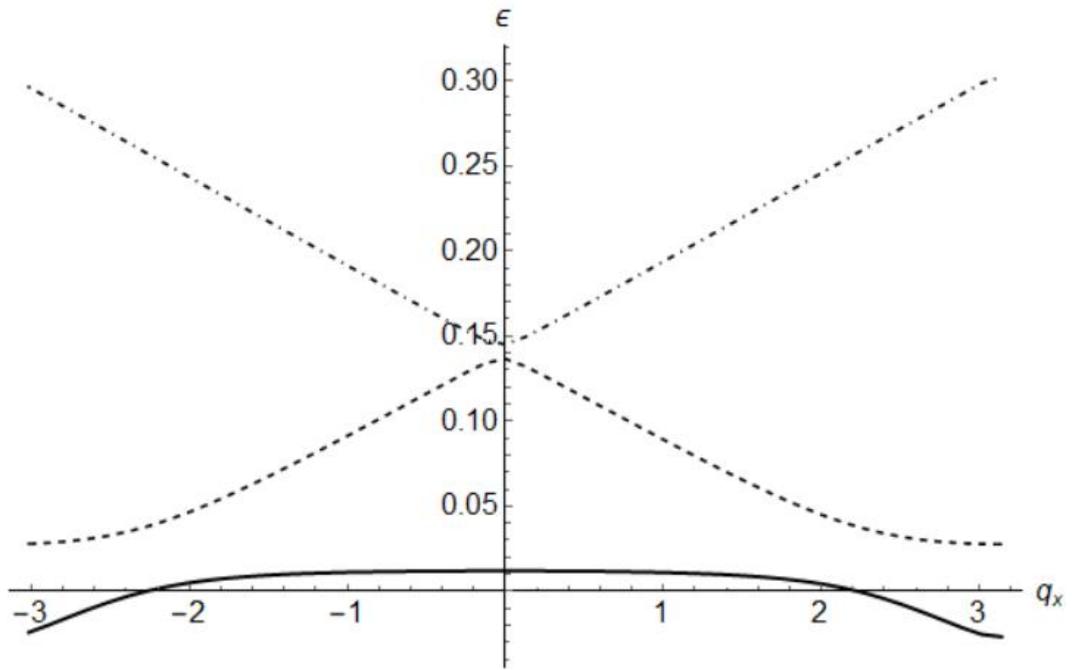

a) $q_y = 0$

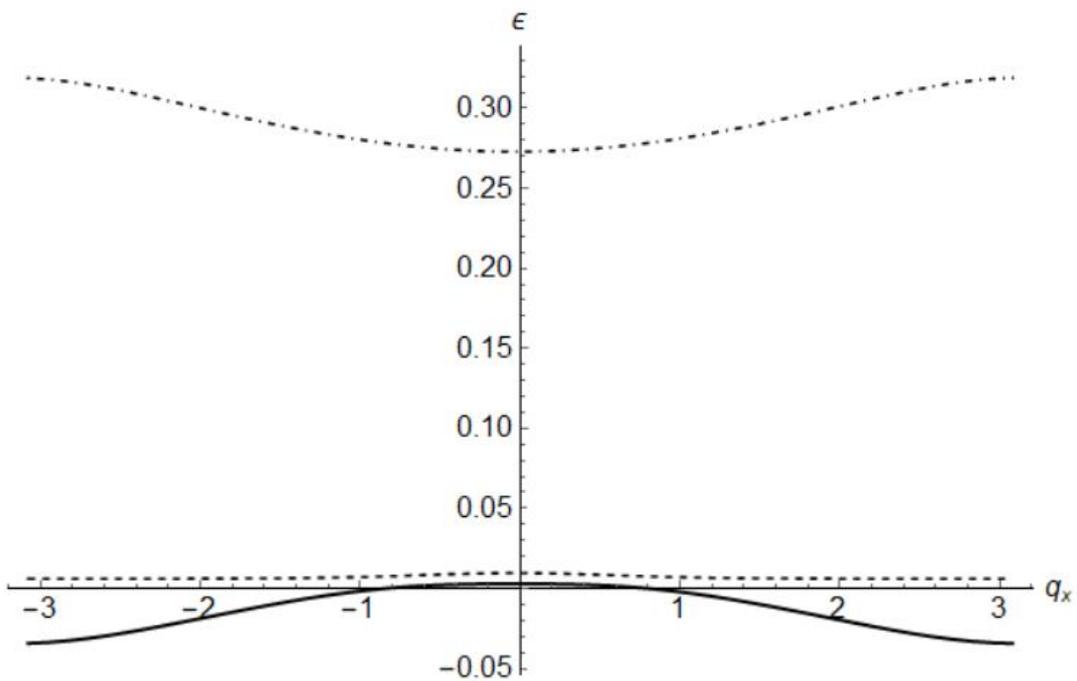

a) $q_y = \pi$

Figure 7. - Dependence of electron energy on the projection of the quasimomentum in the direction of the axis of the superlattice, in minibands closest to the Dirac point in the graphene spectrum in the absence of an external electric field perpendicular to the surface of the sample, in case $d_{II}/d = 3/51$

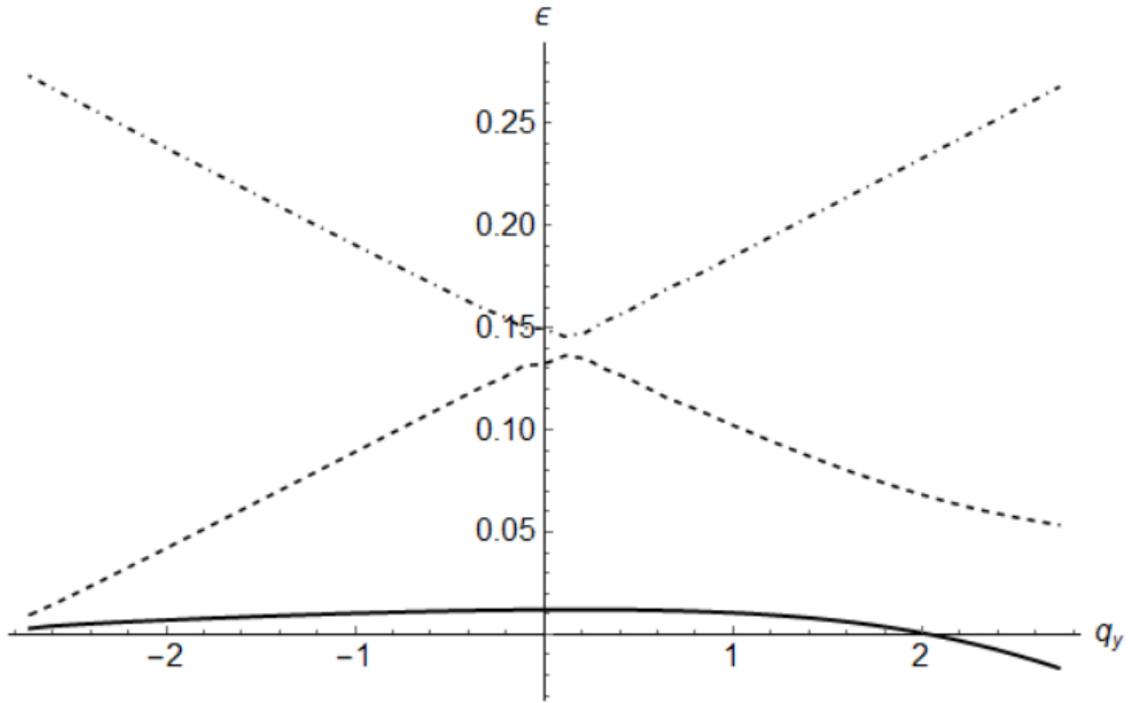

a) $q_x = 0$

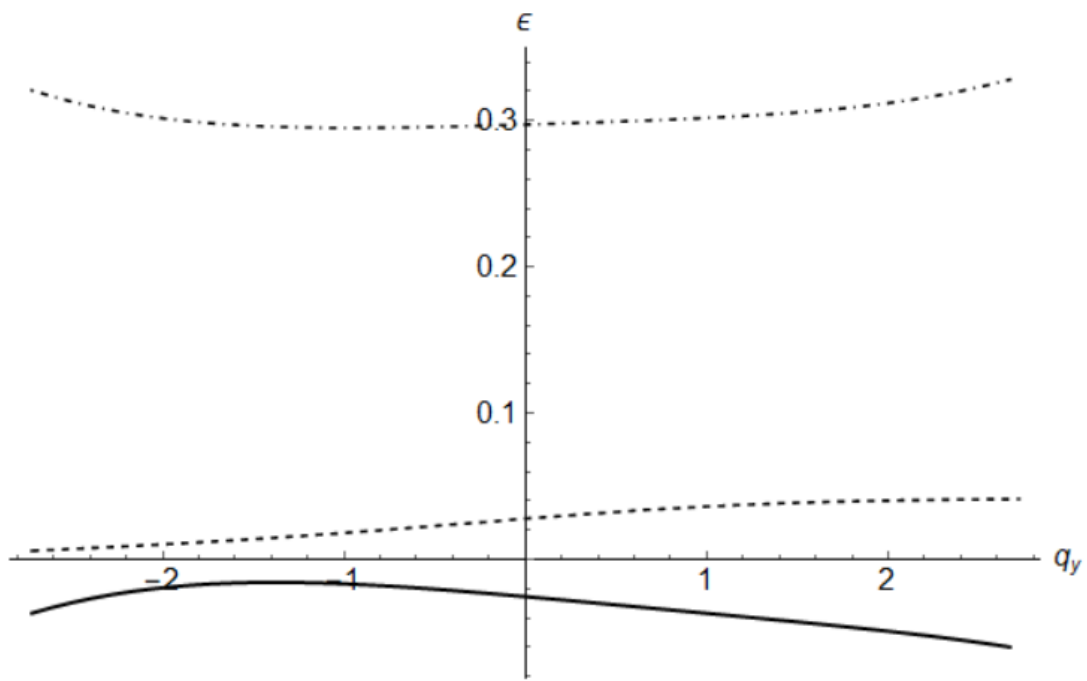

b) $q_x = \pi$

Figure 8. - Dependence of the electron energy on the projection of the quasimomentum in the direction perpendicular to the axis of the superlattice, in minibands closest to the Dirac point in the graphene spectrum, in case $d_{II}/d = 3/51$.

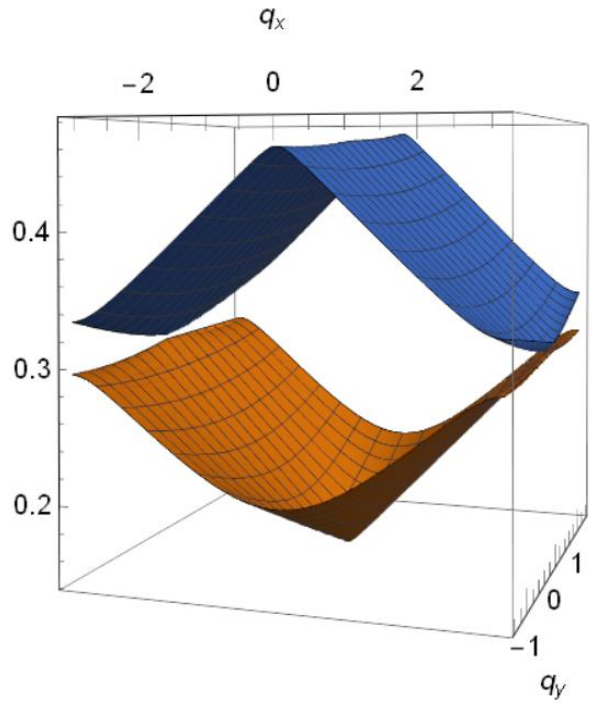

a)

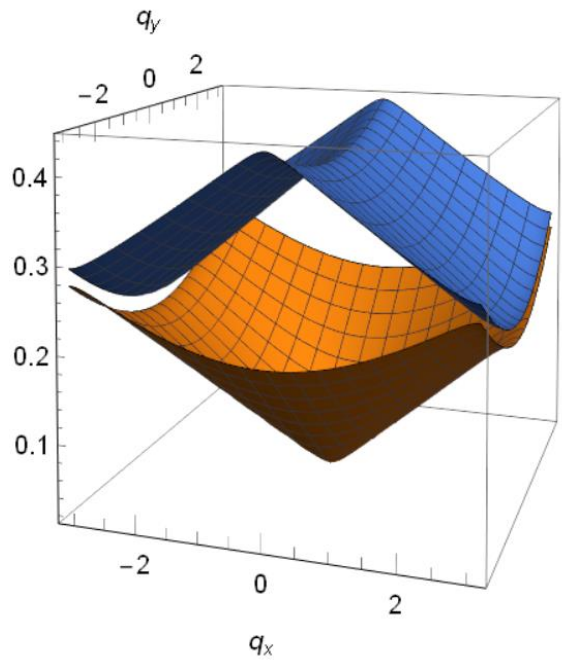

Figure 9. - Dispersion surfaces in case $d_{II} / d = 3/51$ in the absence of an external electric field: a) quantum-chemical calculation; b) calculation based on dispersion equation (22). Energy is measured in electron volts.

Conclusion

Using the analytical approach and quantum-chemical modeling, the electronic states in the GSL, consisting of alternating strips of single-layer and bilayer graphene, placed in a constant electric field perpendicular to the surface of the sample, are investigated. It is shown that the energy spectrum of the structure under consideration has a miniband character, the widths of the forbidden and allowed minibands are hundredths of an electron-volt and can be controlled by an external electric field.

For analytical consideration, the Kronig-Penney model was used. Dispersion equations of two types are obtained, corresponding to two branches in the energy spectrum of bilayer graphene, and each type of dispersion equation also describes the states corresponding to Tamm minibands.

The energy spectrum of the considered GSL was simulated using the methods of the density functional theory implemented in the OpenMX package. Based on quantum-chemical modeling, it was shown that the results of an analytical review regarding the conclusions about the existence of a miniband spectrum symmetric with respect to the K-point of the starting material, single-layer graphene, are valid in the case of superlattices composed of wide strips of single-layer and narrow bands of bilayer graphene. In other cases, the energy spectrum of the SL is not symmetric with respect to the K-point in the direction perpendicular to the SL.

This work was financially supported by grant of the Russian Foundation for Basic Research and the Volgograd Region Administration as part of a scientific project 19-42-343006 r_mol_a.